\documentclass[a4paper]{revtex4}
\usepackage{amsmath}  
\usepackage{amssymb}  
\usepackage{dsfont}   

\begin{document}

\draft
\title{Entanglement for all quantum states}
\author{A. C. de la Torre}\email{delatorre@mdp.edu.ar}
\author{D. Goyeneche}\email{dgoyene@mdp.edu.ar}
\author{L. Leitao}\email{lleitao@mdp.edu.ar} \affiliation{ IFIMAR - (CONICET-UNMDP) \\ Departamento de
F{\'\i}sica
- Universidad Nacional de Mar del Plata\\
Funes 3350, 7600 Mar del Plata, Argentina. \vspace{2cm}}

\begin{abstract}
It is shown that a state that is factorizable in the Hilbert
space corresponding to some choice of degrees of freedom, becomes
entangled for a different choice of degrees of freedom.
Therefore, entanglement is not a special case but is ubiquitous
in quantum systems. Simple examples are calculated and a general
proof is provided. The physical relevance of the change of tensor
product structure is mentioned. \vspace{1cm}
\\ Published in Eur. J. Phys. \textbf{31} (2010) 325-332.

\end{abstract}

\maketitle \vspace{2cm}

\section{INTRODUCTION}

Entanglement is one of the most remarkable features of quantum
mechanics. Consider two exclusive properties of a quantum system,
$A_{1}$ and $A_{2}$, corresponding to two different eigenvalues
of some observable (for instance, spin up and spin down) and also
another unrelated pair of exclusive properties, $B_{1}$ and
$B_{2}$ (for instance, located here or there). Furthermore,
imagine two possible states of the system: $\psi_{1}$,
corresponding to the simultaneous appearance of the properties
$A_{1}$ and $B_{1}$ and the other state, $\psi_{2}$,
corresponding to the appearance of the properties $A_{2}$ and
$B_{2}$. The superposition, $\psi_{1}+\psi_{2}$, is an entangled
state of the system. In this state, none of the properties
$A_{1},A_{2},B_{1},B_{2}$ are objective (in the sense that the
state is \emph{not} an eigenvector corresponding to any of these
eigenvalues) but there are strong quantum correlations among them
because the observation of one property, say $A_{1}$, forces the
appearance of $B_{1}$ although they may be totaly unrelated (like
spin and location). In entangled states all sort of astonishing
quantum effects appear, like violations of Bell's inequalities,
Einstein-Podolsky-Rosen (so called) paradox, Schr{\"o}dinger cat,
nonlocality, contextuality, teleportation, quantum cryptography
and computation, etc. The principle of superposition, that
generates the entanglement, contains perhaps the central essence
of quantum mechanics and almost all pondering concerning the
foundations of quantum mechanics involve entangled states.

The opposite to the entangled states are the factorizable states,
for instance $\psi_{1}$ or $\psi_{2}$ above, where the properties
are objective and the behaviour of the system is closer to
classical expectations; for instance, the correlations found are
understood as a direct consequence of the preparation of the
system. One may erroneously think that there are two classes of
states for the quantum system, entangled and factorizable, that
correspond to qualitative difference in the behaviour of the
system,  close to classical in one case and with strong quantum
correlations in the other. We will see in this work that this is
indeed wrong because factorizable states also exhibit
entanglement with respect to \emph{other} observables. In this
sense, all states are entangled; entanglement is not an
exceptional feature of some states but is ubiquitous in quantum
mechanics.

The fact that factorizability and entanglement are not preserved
in a change of the degrees of freedom used to describe the system
has been analysed by experts, specially those involved in quantum
computation research\cite{Zan,hars}, but this important feature
of quantum mechanics is ignored in textbooks, even advanced ones.
In this work we present simple calculations that emphasize this
remarkable feature and provide thereby a didactic complement for
a modern quantum mechanics course. The calculations mentioned
would be quite involved without the application of the Quantum
Covariance Function that has also received very little
consideration in textbooks. In the following sections we will
define entanglement and factorizability with rigour and we will
prove that every factorizable state becomes entangled in a
different factorization of the Hilbert space. For this, we will
recall a useful tool provided by the Quantum Covariance Function
and we will calculate several explicit examples that may be
useful for teachers and students of quantum mechanics at the
advanced undergraduate and graduate level.

\section{entanglement in compound systems}

The state of a compound quantum system, $S=(S_{1},S_{2})$,
belongs to a Hilbert space build as the tensor product of spaces
corresponding to each subsystem:
$\mathcal{H}=\mathcal{H}_{1}\otimes\mathcal{H}_{2}$. This
decomposition, denoted as a \emph{Tensor Product Structure}
(TPS), may correspond to two individual physical subsystems like,
for instance, one electron and one proton building a hydrogen
atom, or to different degrees of freedom or coordinates of one
system. The degrees of freedom are expected to be independent in
the sense that the assignment of one value to one degree of
freedom is compatible with an arbitrary assignment of \emph{any}
value for the other one. For quantum mechanics, this means that,
in the Hilbert space, the two degrees of freedom $A$ and $B$ will
correspond to two observables whose operators act individually in
each factor space, that is, they are of the form $A\otimes
\mathbb{I}$ and $\mathbb{I}\otimes B$ and therefore they commute.
If we define bases in each factor space,
$\{\varphi_{k}\}\in\mathcal{H}_{1}$ and
$\{\phi_{r}\}\in\mathcal{H}_{2}$, the most general state in the
Hilbert space is given by an expansion in the basis
$\{\varphi_{k}\otimes\phi_{r}\}$ as
\begin{equation}\label{state}
    \Psi=\sum_{k,r} C_{kr}\ \varphi_{k}\otimes\phi_{r}\ .
\end{equation}
This state $\Psi$ is factorizable if there exist $\Psi_{1}
\in\mathcal{H}_{1}$ and $\Psi_{2} \in\mathcal{H}_{2}$ such that
$\Psi=\Psi_{1}\otimes\Psi_{2}$. Otherwise it is entangled. More
precisely, we define entanglement by means of Schmidt
bi-orthogonal decomposition: for each state $\Psi$ there exist
two bases $\{\widetilde{\varphi}_{k}\}\in\mathcal{H}_{1}$ and
$\{\widetilde{\phi}_{r}\}\in\mathcal{H}_{2}$ such that
\begin{equation}\label{biortstate}
    \Psi=\sum_{k=1}^{N} \alpha_{k}\ \widetilde{\varphi}_{k}
    \otimes\widetilde{\phi}_{k}\ ,
\end{equation}
where $N\leq\min\{D_{1},D_{2}\}$ and where $D_{1}$ and $D_{2}$
are the dimensions of the Hilbert spaces. Notice that in this
expansion we do not have a double sum as in the general expansion
in Eq.(\ref{state}). The bases for the bi-orthogonal
decomposition are not unique and, of course, depend on the state
$\Psi$. If $N=1$ the state is factorizable and if $N\geq 2$ the
state is entangled.

Notice that in the determination of whether a state is
factorizable or entangled, the factorization of the Hilbert space
(that is, the TPS) is crucial and this factorization depends on
the choice of the observables corresponding to the degrees of
freedom. From the mathematical point of view, every TPS is
equivalent but from the physical point of view, the TPS are
determined operationally by the measurements and operations that
are accessible under given physical circumstances. For instance,
if our composite system consists of two particles that are
spatially separated, the most natural TPS is given by the tensor
product of the single-particle Hilbert spaces associated with the
individual particles. However if in this same system, the overall
motion is uninteresting and only the relative  motion is relevant
we may prefer a TPS corresponding, not to the position of the
individual particles, but instead, to the center of mass and
relative position.

A relevant question is whether some arbitrary state of a system,
analysed with different choices of the degrees of freedom, that
is, with different TPS, still maintains the property of been
factorizable or not. In other words, is factorizability an
objective property of the system or is it a feature of our
description of the system. In order to approach this question we
recall a useful tool provided by the Quantum Covariance Function
that relates a state $\Psi$ and two observables $A$ and $B$.

\section{the quantum covariance function}

Given two arbitrary hermitian operators $A$ and $B$ and a
normalized Hilbert space element $\Psi$, we define the
\emph{Quantum Covariance Function} (QCF) by
\begin{equation}\label{QCF}
 Q(A,B,\Psi)=\langle \Psi,AB\Psi\rangle - \langle
\Psi,A\Psi\rangle\langle\Psi,B\Psi\rangle\ .
\end{equation}
This function has been studied in detail\cite{dlt1} and was use
to provide an elegant proof of the uncertainty principle in the
most general form given by Schr{\"o}dinger\cite{schro}. Another
interesting application of this function allows a complete study
of manifest and concealed correlations in quantum
systems\cite{dlt2}. For our purpose here, we notice that when the
observables correspond to two degrees of freedom, that is, they
are of the form $A\otimes \mathbb{I} $ and $\mathbb{I}\otimes  B
$, and the state is factorizable, $\Psi=\Psi_{1}\otimes\Psi_{2}$,
then the QCF vanishes. The proof is trivial: $Q(A\otimes
\mathbb{I}\ ,\ \mathbb{I}\otimes  B,\ \Psi_{1}\otimes\Psi_{2})=
\langle \Psi_{1}\otimes\Psi_{2}\ ,\ A\otimes B\
\Psi_{1}\otimes\Psi_{2}\rangle - \langle\Psi_{1}\ ,
A\Psi_{1}\rangle\|\Psi_{2}\|^{2}\|\Psi_{1}\|^{2}\langle\Psi_{2}
,B\Psi_{2}\rangle\ = 0\ .$ Notice that $\Psi$ factorizable
implies $Q(A,B,\Psi)=0$ but the inverse is not true: there are
cases with vanishing QCF but with entangled states as can be seen
in ref.\cite{dlt2}. In any case, if the QCF does not vanish, then
we are sure that the state is not factorizable, that is, it is
entangled.

\section{transformation entanglement }

Let us consider a quantum system  with two subsystems
$S=(S_{A},S_{B})$ that may correspond to two degrees of freedom
$A$ and $B$. The state of the system belongs then to the Hilbert
space $\mathcal{H}=\mathcal{H}_{A}\otimes\mathcal{H}_{B}$ and the
two degrees of freedom are represented by operators $A\otimes
\mathbb{I} $ and $\mathbb{I}\otimes  B $. Let us consider a
factorizable, non entangled, state $\Psi=\Psi_{A}\otimes\Psi_{B}$
with $\Psi_{A}$ and $\Psi_{B}$ arbitrary states (not necessarily
eigenvectors of $A$ and $B$) in the spaces $\mathcal{H}_{A}$ and
$\mathcal{H}_{B}$. Then there exists a transformation of the
degrees of freedom $F=F(A,B)$ and $G=G(A,B)$ that suggests a
different factorization or TPS,
$\mathcal{H}=\mathcal{H}_{F}\otimes\mathcal{H}_{G}$, where the
state is no longer factorizable:
$\Psi\neq\Psi_{F}\otimes\Psi_{G}$ with
$\Psi_{F}\in\mathcal{H}_{F}$ and $\Psi_{G}\in\mathcal{H}_{G}$.
The state becomes entangled in these new degrees of freedom; the
factorizability of \emph{states} is not invariant under a
different factorization of the Hilbert space.

We will next clarify this with two simple examples and we will
later give a general proof.

\subsection{ System with two coordinates }

Let us consider a very simple system characterized by two space
coordinates $X_{1}$ and $X_{2}$. This may correspond to the
position of two free particles moving in a line or to one
particle moving in a plane (the equivalence of $n$ free particles
moving in $\mathbb{R}_{1}$ with one particle moving in
$\mathbb{R}_{n}$ is an interesting fact that has been used to
relate nonseparability with contextuality\cite{dlt3}). The state
of the system is then an element of
$\mathcal{H}_{1}\otimes\mathcal{H}_{2}$ and the two coordinates
correspond to the operators $ X_{1}=X\otimes \mathbb{I} $ and $
X_{2}=\mathbb{I}\otimes X  $. If we use the basis
$\{\varphi_{x}\}\in\mathcal{H}_{1}$ and
$\{\phi_{x}\}\in\mathcal{H}_{2}$ corresponding to the
eigenvectors of the position operator in both spaces, then the
most general factorizable state is given by
\begin{equation}\label{estfac}
 \Psi =  \left(\int\!\!\! dx_{1}\  f(x_{1})\varphi_{x_{1}}\right)\otimes
 \left(\int\!\!\! dx_{2}\  g(x_{2})\phi_{x_{2}}\right)
 =\int\!\!\! dx_{1}\  \int\!\!\! dx_{2}\  f(x_{1})g(x_{2})\ \varphi_{x_{1}}\otimes
\phi_{x_{2}}\ ,
\end{equation}
where $f(x)$ and $g(x)$ are two properly normalized arbitrary
functions. (For more rigour, we should mention that the spaces $
\mathcal{H}_{1}$ and $ \mathcal{H}_{2}$ are \emph{rigged} Hilbert
spaces that contain not only the normalizable square integrable
functions but also the improper eigenvectors $\{\varphi_{x}\}$
and $\{\phi_{x}\}$. More details on this can be found in advanced
quantum mechanics books\cite{ball}).

Instead of $X_{1}$ and $X_{2}$ we can now consider another pair
of degrees of freedom given by
\begin{eqnarray}
  A &=& X_{1}+X_{2} =  X\otimes\mathbb{I} +  \mathbb{I}\otimes X  \\
  B &=&  X_{1}-X_{2} = X\otimes\mathbb{I} - \mathbb{I}\otimes X \
  .
\end{eqnarray}
Physically, these new coordinates are related to the center of
mass and relative distance, in the case of two particles in a
line, or to a rotation and reflection of the coordinate axis in
the case of one particle moving in a plane. It is a trivial
change of variables but, as we will see, with significant
consequences for the treatment of the quantum system. Let
$\chi_{a,b}$ in $\mathcal{H}_{1}\otimes\mathcal{H}_{2}$ denote
the eigenvectors of $A$ and $B$ corresponding to the eigenvalues
$a$ and $b$. That is, $A\chi_{a,b}=a\chi_{a,b}$ and
$B\chi_{a,b}=b\chi_{a,b}$. One can easily check that
$\chi_{a,b}=\varphi_{\frac{1}{2}(a+b)}\otimes
\phi_{\frac{1}{2}(a-b)}$. Consider now the degree of freedom $A$
alone, isolated from the other degree of freedom $B$. To this
degree of freedom we can associate a Hilbert space
$\mathcal{H}_{A}$ spanned by the eigenvectors of $A$,
$\{\eta_{a}\}$. Similarly, the eigenvectors $\{\xi_{b}\}$ of $B$,
considered independently, generate another Hilbert space
$\mathcal{H}_{B}$. The tensor product of these spaces,
$\mathcal{H}_{A}\otimes\mathcal{H}_{B}$, provide a different
factorization of the Hilbert space of the compound system,
spanned by the basis $\{\eta_{a}\otimes \xi_{b}\}$. The two bases
 $\{\eta_{a}\otimes \xi_{b}\}$ and $\{\varphi_{x_{1}}\otimes
\phi_{x_{2}}\}$ are related by
\begin{eqnarray}
  \eta_{a}\otimes \xi_{b} &=&  \varphi_{\frac{1}{2}(a+b)}\otimes
\phi_{\frac{1}{2}(a-b)}\\
\varphi_{x_{1}}\otimes \phi_{x_{2}}&=& \eta_{x_{1}+x_{2}}\otimes
\xi_{x_{1}-x_{2}}\ .
\end{eqnarray}
Notice that this change of basis is trivial, it amounts only to a
relabelling or reordering of the basis elements. This is expected
because the new and old degrees of freedom commute and therefore
they share the same basis. Anyway, if we perform the variable
change $x_{1}+x_{2}=a,\ x_{1}-x_{2}=b, \
dx_{1}dx_{2}=\frac{1}{2}dadb$ in the factorizable state in
Eq.(\ref{estfac}) we get the same state in the new basis given as
\begin{equation}\label{estnofac}
 \Psi = \frac{1}{2}\int\!\!\! \int\!\!\!
 da\   db\  f\left(\frac{1}{2}(a+b)\right)g\left(\frac{1}{2}(a-b)\right)\
 \eta_{a}\otimes \xi_{b}\ .
\end{equation}
One can, of course, find many examples of factorizable states
that remain factorizable after the change of variables. For
instance, the gaussian
$f(x_{1})g(x_{2})=\exp[-x^{2}_{1}-x^{2}_{2}]$ becomes
$\exp[(-a^{2}-b^{2})/2]$ that is also factorizable, or also a
plane wave $\exp[i(k_{1}x_{1} + k_{2}x_{2}]$ remains
factorizable. However, not \emph{every} factorizable state
remains so and, in general,
$f(\frac{1}{2}(a+b))g(\frac{1}{2}(a-b))\neq F(a)G(b)$. As an
example for nonfactorizability we can take, for instance, a
double gaussian for
$f(x_{1})=\exp(-(x_{1}-d)^{2})+\exp(-(x_{1}+d)^{2})$ and one
single gaussian for $g(x_{2})=\exp(-x_{2}^{2}) $.

A more elegant proof that the variable change destroys the
factorizability is obtained using the QCF. Clearly, with the
factorizable state in Eq.(\ref{estfac}), that is with
$\Psi=\Psi_{1}\otimes\Psi_{2}$, the QCF vanishes:
\begin{eqnarray}
\nonumber Q(X_{1},X_{2},\Psi) &=& \langle\Psi,X\otimes
X\Psi\rangle- \langle\Psi ,X\otimes
\mathbb{I}\Psi\rangle\langle\Psi,\mathbb{I}\otimes X\Psi\rangle \\
 &=&\langle\Psi_{1},X\Psi_{1}\rangle\langle\Psi_{2},X\Psi_{2}\rangle-
\langle\Psi_{1},X\Psi_{1}\rangle\|\Psi_{2}\|^{2}
\|\Psi_{1}\|^{2}\langle\Psi_{2},X\Psi_{2}\rangle = 0\ .
\end{eqnarray}
However for the same state, a similar calculation gives
\begin{eqnarray}
\nonumber Q(A,B,\Psi) &=&
\langle\Psi_{1},X^{2}\Psi_{1}\rangle-\langle\Psi_{2},X^{2}\Psi_{2}\rangle-
\langle\Psi_{1},X\Psi_{1}\rangle^{2}+
\langle\Psi_{2},X\Psi_{2}\rangle^{2} \\ &=&
\Delta^{2}_{X_{1}}-\Delta^{2}_{X_{2}} \neq 0
\end{eqnarray}
because, in general, the ``widths'' of $|f(x_{1})|^{2}$ and
$|g(x_{2})|^{2}$ are different.

Summarizing, a factorizable state in the compound Hilbert space
corresponding to the degrees of freedom $X_{1}$ and $X_{2}$
becomes entangled when we consider the degrees of freedom
$A=X_{1}+X_{2}$ and $B=X_{1}-X_{2}$ and viceversa. Instead of
this simple transformation of the degrees of freedom we can
consider any general reversible map between the coordinates
$(X_{1},X_{2})$ and $(A,B)$. In this case, the state in
Eq.(\ref{estfac}) becomes
\begin{equation}\label{estnofacGen}
 \Psi = \int\!\!\! \int\!\!\!
da\   db\  \left| \frac{\partial (x_{1},x_{2})}{\partial
(a,b)}\right| \
f\left(\frac{}{}x_{1}(a,b)\right)g\left(\frac{}{}x_{2}(a,b)\right)\
 \eta_{a}\otimes \xi_{b}\ .
\end{equation}
We can prove that, in general, this expression is not
factorizable for arbitrary normalized functions $f(x_{1})$ and
$g(x_{2})$. In fact, if there exist functions $F(a)$ and $G(b)$
not vanishing everywhere such that
\begin{equation}\label{factFG}
f\left(\frac{}{}x_{1}(a,b)\right)g\left(\frac{}{}
x_{2}(a,b)\right) = F(a) G(b)\ ,
\end{equation}
then we reach a contradiction. In order to see this, let us
choose, among all possible $f(x_{1})$, one that has a zero in
$x_{0}$, that is, $f(x_{0})=0$. Therefore in the $(x_{1},x_{2})$
plane, the product $f\left(x_{1}\right)g\left(x_{2}\right)$
vanishes along a straight line perpendicular to the $x_{1}$ axis.
This line is mapped in the $(a,b)$ plane into a curve given by $
x_{1}(a,b)=x_{0}$. We can solve this equation for $a$, that is,
$a=a_{0}(b)$ and replace it in the right hand side of
Eq.(\ref{factFG}) obtaining $F(a_{0}(b))G(b)=0\ \forall b$. Now,
since $\forall b,\ G(b)\neq 0$ by definition, we must have
$F(a_{0}(b))=0\ \forall b$, in contradiction with the assumption
that $F$ does not vanishes everywhere.

From the example seen, it becomes clear that every change to new
degrees of freedom that mixes the old ones, destroys the
factorization of the state. The only possibility to preserve the
factorization is when  the degrees of freedom are not mixed, that
is, the transformations are of the type $A=A(X_{1})$ and
$B=B(X_{2})$. Therefore, the unitary transformations in the
Hilbert space $\mathcal{H}_{1}\otimes\mathcal{H}_{2}$ that leave
the TPS invariant are of the type $U_{A}\otimes U_{B}$, for
instance, the time evolution of internal and external degrees of
freedom\cite{hars}. A generalization of this to TPS involving any
number of factors is evident.

\subsection{ System with two spins }

In this example, let us consider a system of two particles with
spin 1/2. As degrees of freedom to characterize the system we can
take, as is usually done, the $z$ component of both spins
$S_{1z}$ and $S_{2z}$. Their corresponding two dimensional
Hilbert spaces $\mathcal{H}_{1}$ and $\mathcal{H}_{2}$ are
spanned by the two basis $\{\varphi_{\pm}\}$ and $\{\phi_{\pm}\}$. The
four dimensional Hilbert space
$\mathcal{H}=\mathcal{H}_{1}\otimes\mathcal{H}_{2}$ for the
system has a basis $\{\psi_{k,r}=\varphi_{k}\otimes\phi_{r},\
k,r=+,-\}$ and the most general factorizable state is given by
\begin{equation}\label{fact}
\Psi=\Psi_{1}\otimes\Psi_{2}=
\left(\sum_{k}\alpha_{k}\varphi_{k}\right)\otimes
\left(\sum_{r}\beta_{r}\phi_{r}\right)
=\sum_{k,r}\alpha_{k}\beta_{r}\psi_{k,r} \ .
\end{equation}
We will see that when we factorize the Hilbert space
corresponding to other  degrees of freedom this state becomes
entangled.

As different degrees of freedom we can take, for instance, the
square of two orthogonal components of total spin.
\begin{eqnarray}
S_{z}^{2} &=& \left(S_{z}\otimes\mathbb{I}+\mathbb{I}\otimes
S_{z}\right)^{2} =\frac{\hbar^{2}}{2}
\mathbb{I}\otimes\mathbb{I}+ 2\ S_{z}\otimes S_{z}
   \\
S_{x}^{2} &=& \left(S_{x}\otimes\mathbb{I}+\mathbb{I}\otimes
S_{x}\right)^{2} =\frac{\hbar^{2}}{2}
\mathbb{I}\otimes\mathbb{I}+ 2\ S_{x}\otimes S_{x} \ .
\end{eqnarray}
These two operators commute $[S_{z}^{2},S_{x}^{2}]=0$ and their
eigenvalues are $0$ and $\hbar^{2}$ corresponding to the
degeneracy eigenvectors $\{\chi_{s,t},\ s,t=0,1\}$. That is,
\begin{eqnarray}
\nonumber
  S_{z}^{2}\chi_{s,t} &=& s\hbar^{2}\chi_{s,t}
   \\
 S_{x}^{2}\chi_{s,t} &=& t\hbar^{2}\chi_{s,t} ,\ s,t= 0,1\ .
\end{eqnarray}
We have then two different bases $\{\psi_{k,r},\ k,r=+,-\}$ and
$\{\chi_{s,t},\ s,t=0,1\}$. In this example, the new and old
degrees of freedom do not commute and they do not share the basis
as was the case in the previous example. The unitary
transformation relating both bases is
\begin{equation}\label{unittrans}
 \left(%
\begin{array}{c}
  \chi_{1,1} \\
  \chi_{1,0} \\
  \chi_{0,1} \\
  \chi_{0,0} \\
\end{array}%
\right) =\frac{1}{\sqrt{2}}
\left(%
\begin{array}{cccc}
  1 & 1 & 0 & 0 \\
  1 & -1 & 0 & 0\\
 0 & 0 & 1 &1 \\
  0 &0 & 1 & -1 \\
\end{array}%
\right)
\left(%
\begin{array}{c}
  \psi_{+,+} \\
  \psi_{-,-}\\
  \psi_{+,-}\\
  \psi_{-,+}\\
\end{array}%
\right)\ .
\end{equation}
Now, if we take the most general factorizable state in the TPS
related with the basis $\{\psi_{k,r} \}$, given in
Eq.(\ref{fact}), and we make the change of basis with the unitary
transformation, then one can prove that the resulting expression
is, in general, not factorizable in the TPS related to
$\{\chi_{s,t} \}$, that is, it can not always be written in the
form
\begin{equation}\label{nofact}
\Psi = \sum_{s,t}\delta_{s}\eta_{t}\chi_{s,t} \ .
\end{equation}
The change of basis destroys the factorizability. Instead of this
explicit long calculation it is simpler to use the QCF to show
that the state is entangled in the Hilbert space factorization
corresponding to the new variables. Doing this we obtain
\begin{equation}\label{qcfspin}
 Q(S_{z}^{2},S_{x}^{2},\Psi)=
-\hbar^{2}\langle\Psi_{1},S_{y}\Psi_{1}\rangle
\langle\Psi_{2},S_{y}\Psi_{2}\rangle - 4\
\langle\Psi_{1},S_{x}\Psi_{1}\rangle
\langle\Psi_{2},S_{x}\Psi_{2}\rangle
\langle\Psi_{1},S_{z}\Psi_{1}\rangle
\langle\Psi_{2},S_{z}\Psi_{2}\rangle
 \neq 0\
\end{equation}
for arbitrary $\Psi_{1}$ and $\Psi_{2}$.

\subsection{ General proof }

The simple example concerning two spatial coordinates can be
generalized to provide a general proof. Let us consider a quantum
system  with two subsystems $S=(S_{A},S_{B})$ described in the
Hilbert space $\mathcal{H}=\mathcal{H}_{A}\otimes\mathcal{H}_{B}$
corresponding to the TPS associated with the two degrees of
freedom $A\otimes \mathbb{I} $ and $\mathbb{I}\otimes  B $. Let
us assume a factorizable, non entangled, normalized state
$\Psi=\Psi_{A}\otimes\Psi_{B}$ where $\Psi_{A}$ and $\Psi_{B}$
are two arbitrary normalized states in the factor spaces
$\mathcal{H}_{A}$ and $\mathcal{H}_{B}$. Then there exists a
transformation of the degrees of freedom, $F=A\otimes \mathbb{I}
+\mathbb{I}\otimes B $ and $G=A\otimes \mathbb{I}
-\mathbb{I}\otimes  B$, whose TPS has a different factorization,
$\mathcal{H}=\mathcal{H}_{F}\otimes\mathcal{H}_{G}$, where the
state is no longer factorizable. In order to prove this we show
that the QCF $Q(F,G,\Psi)\neq 0$ and therefore the state $\Psi$
is entangled in the Hilbert space factorization
$\mathcal{H}=\mathcal{H}_{F}\otimes\mathcal{H}_{G}$ corresponding
to the degrees of freedom $F$ and $G$. Notice that $FG=(A\otimes
\mathbb{I} +\mathbb{I}\otimes  B )(A\otimes \mathbb{I}
-\mathbb{I}\otimes  B)=A^{2}\otimes \mathbb{I} -\mathbb{I}\otimes
B^{2}$ and then we have
\begin{eqnarray}
\nonumber Q(F,G,\Psi)&=& \langle\Psi_{A}\otimes\Psi_{B}\ ,\
(A^{2}\otimes \mathbb{I} -\mathbb{I}\otimes
B^{2})\Psi_{A}\otimes\Psi_{B}\rangle \ -\\
\nonumber& &
 \langle\Psi_{A}\otimes\Psi_{B}\ ,\ (A\otimes \mathbb{I}
+\mathbb{I}\otimes  B  )\Psi_{A}\otimes\Psi_{B}\rangle
 \langle\Psi_{A}\otimes\Psi_{B}\ ,\ (A\otimes \mathbb{I}
-\mathbb{I}\otimes  B )\Psi_{A}\otimes\Psi_{B}\rangle\\
\nonumber
  &=&\langle\Psi_{A},A^{2}\Psi_{A}\rangle -
   \langle\Psi_{B},B^{2}\Psi_{B}\rangle  -
 \langle\Psi_{A},A \Psi_{A} \rangle^{2}+
 \langle\Psi_{B},B \Psi_{B}\rangle ^{2} \\
   &=& \Delta^{2}_{A}-\Delta^{2}_{B}\neq 0\
\end{eqnarray}
because the indeterminacies of $A$ and $B$ in the arbitrary
states $\Psi_{A}$ and $\Psi_{B}$ are, in general, different.

\section{ conclusions }

We have seen that the factorizability of a state is a property
that is not invariant under a change of the degrees of freedom
that we use in order to describe the system. This proof is made
simple by the use of the QCF whose non-vanishing is a criterium
for entanglement.

The fact that the appearance of entanglement depends on the
choice of degrees of freedom can find an interesting application
in the ``disentanglement'' of a state. One can, sometimes,
transform an entangled state into a factorizable one by a
judicious choice of the degrees of freedom. In some sense this is
the inverse problem to the one presented in section IV. One
example of this is provided in Ref.\cite{referee} where the
entangled state of the compound system of one proton and one
electron with Coulomb interaction becomes factorizable when we
use center of mass and relative position coordinates instead of
the individual spacial coordinates of the proton and electron. In
this case, the two particle system consisting in one proton and
one electron in an entangled state is described in a simpler,
factorizable, state of two fictitious noninteracting particles:
one with the total mass of the hydrogen atom, moving freely in
space, and another particle with the effective mass moving in a
fixed Coulomb potential.

Perhaps the most important manifestation of quantum correlations,
that is, those that can not be explained in terms of some
classical interaction, involves the violations of Bell's
inequalities. Furthermore, it has been shown\cite{gis} that in
every nonfactorizable or entangled state there are observables
that violate Bell's inequalities. In this work, we have seen that
for any system in a factorizable state, we can find different
degrees of freedom that suggest a different factorization of the
Hilbert space where the same state becomes entangled. As a
consequence of this we can conclude that in every state, even for
those factorizable, we can find pairs of observables that will
violate Bell's inequalities. This violation of the classical
behaviour is then not exceptional but is ubiquitous in quantum
systems.

\end{document}